\documentclass[12pt]{article}
\usepackage[dvips]{graphicx}
\usepackage{epsfig}

\renewcommand{\bar}[1]{\overline{#1}}

\newcommand{\bit}[1]{\mbox{\boldmath$#1$}}

\begin{document}
\begin{flushright}
{\small
SLAC--PUB--12062\\
UK/TP-06-08 \\ 
August 2006 \\}
\end{flushright}

%sjb 8-17-06

\begin{center}
{\Large
\bf 
Evidence for the Absence of Gluon Orbital Angular Momentum in the Nucleon
}\footnote{Work 
supported in part by the U.S. Department of Energy under 
contracts DE--AC02--76SF00515 and DE--FG02--96ER40989
} \vspace{1.5cm}
%DE-FG02-96ER40989 and DE-FG01-00ER45832.

\bigskip

Stanley J. Brodsky\\
Stanford Linear Accelerator Center, Stanford University, Stanford,
California 94309

\bigskip

Susan Gardner\\
Department of Physics and Astronomy, University of Kentucky,
Lexington, Kentucky 40506-0055

\bigskip
\date{\today}
\bigskip
\end{center}
\bigskip

{\noindent 
The Sivers mechanism for the
single-spin asymmetry in unpolarized lepton scattering from a transversely
polarized nucleon
is driven by the orbital angular momentum carried
by its quark and gluon constituents, combined with QCD final-state
interactions. Both quark and gluon mechanisms can generate
such a single-spin asymmetry, though only the
quark mechanism can explain the small single-spin asymmetry
measured by the COMPASS collaboration
on the deuteron, suggesting the gluon mechanism is
small relative to the quark mechanism.
We detail empirical studies through which the
gluon and quark orbital angular momentum contributions, quark-flavor by
quark-flavor, can be elucidated.
}

\vfill\vfill
\newpage

The nucleon is a composite particle with spin $1/2$. There is little
doubt that the theory of quantum chromodynamics (QCD) describes the manner
in which the nucleon's spin is carried by its constituents, yet 
clarifying the details of this picture has incited intense
theoretical and experimental activity~\cite{Filippone:2001ux}. 
Much has been made of the 
empirical fact that the spin of the nucleon is not given 
by the net helicity of its valence quarks~\cite{Ashman:1987hv};
however, this  is not so much a ``crisis'' for QCD 
as it is for the non-relativistic quark model, since
the latter rationalizes the charges, spins, and magnetic moments
of the baryons in terms of the properties of its constituent quarks.
The rich structures revealed through 
deeply inelastic scattering experiments on the proton~\cite{cteq}
and through Drell-Yan production of massive lepton 
pairs with hadron beams and targets~\cite{Hawker:1998ty}, 
compellingly demonstrate the limitations of such a simple picture.
% --- 
%not only the existence of the 
%gluon and sea-quark constituents, 
%but also the important role of internal orbital angular momentum. 

The nucleon contains both quark and gluon components in QCD, so that 
its spin of $1/2$ must follow from the sum of the spin 
and orbital angular momenta carried by these constituents: 
\begin{equation}
\frac{1}{2} = L_q^{\rm net}  + \frac{1}{2} \Delta \Sigma + L_g + \Delta g \,,
\end{equation}
where we write $\Delta \Sigma$ for the net helicity of the quarks. 
Our decomposition
is referenced to a polarization axis, so that $L_q^{\rm net}$ and 
$L_g$ are the components
of the orbital angular momentum, due to quark and gluon constituents, 
respectively, with
respect to that axis. 
The decomposition is not unique~\cite{Jaffe:1989jz}. 
In what follows, we will use the decomposition 
based on 
the angular momentum tensor 
in light-cone gauge, $A^+=0$~\cite{Jaffe:1989jz}, 
so that the gluon constituents have physical polarization $S^z = \pm 1$, 
and there are no ghosts~\cite{BPPrev}.
Our conclusions concerning the decomposition of the nucleon spin 
will thus be specific to light-cone gauge, giving us a natural
connection to the physics of light-front wave functions~\cite{BPPrev}, which
are invariably 
defined in this gauge. 
We note that 
a manifestly gauge-invariant decomposition is also possible~\cite{Ji:1996ek}. 

Light-front wave functions (LFWFs) enjoy 
many advantages: they are frame independent, and the spins of the
constituents satisfy $J^z = \sum_{i=1}^n S_i^z + \sum_{i=1}^{n-1} L_i^z$
Fock state by Fock state for a  polarization axis $\mathbf{z}$ 
--- we emphasize that 
there are only $n-1$ internal orbital
angular momenta for a given Fock state with $n$ constituents.  The LFWFs 
are the eigensolutions 
of the QCD Hamiltonian defined at fixed light-front time $\tau = t + z/ c.$
Indeed, LFWFs
are the natural
way to understand the structure of hadrons
as probed through lepton scattering experiments.  
For example, 
the computation of the 
electromagnetic elastic 
form factors in the light-front formalism~\cite{DrellYan,West}
yields the insight~\cite{BrodskyDrell} that 
the anomalous magnetic moment $\kappa \equiv (e/2M)F_2(0)$
is non-zero only if the quark Fock components carry non-zero transverse 
orbital angular momentum, i.e., if $
\mathbf{S}_\perp\cdot \mathbf{L}_{\perp}^q \ne 0$. 
We neglect fundamental $T$ violation, so that 
$\kappa$ is real~\cite{Brodsky:2006ez}. 
Since the proton's anomalous magnetic 
moment is nearly twice that of its Dirac magnetic moment, a ``spin crisis'' 
in DIS could have been 
altogether expected.

In this paper we will study constraints on the orbital angular momentum of the nucleon's constituents
using the azimuthal 
single-spin asymmetries produced from a target polarized 
transverse to the reaction plane. 
Such asymmetries have been seen in a 
variety of reactions, although we focus on those observed
in semi-inclusive deeply inelastic lepton scattering (SIDIS), as in, e.g., 
$\ell p^\uparrow \to \ell^\prime \pi^\pm X$. 
In general, the single-spin asymmetry is proportional to the invariant form
$\epsilon_{\mu \nu \sigma \tau} P^\mu S_p^\nu p_\pi^\sigma q^\tau$ 
where $S_p$, the nucleon spin, satisfies $S_p^2 =-1$ and 
$S_p \cdot P=0$, where $P$ is the nucleon momentum and $q=p_{l^\prime} - p_{l}$
is the momentum transfer, with $Q^2=-q^2$. 
The correlation is proportional to 
$\mathbf{S}_p\cdot \mathbf{p}_\pi \times \mathbf{q}$ in the
target rest frame, and since it is of leading twist, it obeys
Bjorken scaling~\cite{Brodsky:2002cx}. 
This pseudo-T-odd correlation is engendered
by final-state interactions (FSI) of the struck quark, 
and thus it does not reflect a fundamental violation
of time-reversal invariance~\cite{Brodsky:2002cx,collinsT}. 
The discrete symmetry transformations in the light-front formalism 
are studied in detail in Ref.~\cite{Brodsky:2006ez}. 

The azimuthal single-spin asymmetry (SSA) 
for $\pi^\pm$ production in SIDIS from a unpolarized beam
and transversely polarized target, is defined as~\cite{Airapetian:2004tw}
\begin{eqnarray}
A^{\pi^\pm}_{UT} (\phi, \phi_s) 
&\equiv& \frac{1}{|\langle S_p \rangle|}
\left(
\frac{N_{\pi^\pm}^\uparrow (\phi, \phi_s) - N_{\pi^\pm}^\downarrow (\phi, \phi_s)}
{N_{\pi^\pm}^\uparrow (\phi, \phi_s) + N_{\pi^\pm}^\downarrow (\phi, \phi_s)}
\right) \,, \nonumber\\
&\equiv& A_{UT}^C \sin(\phi + \phi_S)  + A_{UT}^S \sin(\phi - \phi_S) + ....\,,
\label{autdef}
\end{eqnarray} 
where 
the ``C'' and ``S'' superscripts refer to the asymmetries generated by the 
Collins and Sivers effects, respectively, and 
the definition of the
azimuthal angles $\phi$ and $\phi_S$ are as in 
Ref.~\cite{Airapetian:2004tw}. 
In the Collins mechanism~\cite{Collins}, the
asymmetry is formed through the product of the transversity
distribution and a pseudo-T-odd spin-dependent fragmentation function 
which describes
the correlation of the transverse polarization of the struck quark
with the transverse momentum of the produced hadron. In the Sivers
mechanism~\cite{Sivers}, the asymmetry arises from the product of
a pseudo-T-odd distribution function 
which describes
the correlation of the transverse momentum of the struck quark with
the transverse nucleon spin 
and a spin-independent
fragmentation function~\cite{Anselmino:1994tv,Boer:1997nt}. 
The asymmetries arising from the 
two mechanisms differ in their 
precise azimuthal angle dependence and can be separated 
as in Eq.~(\ref{autdef})~\cite{Airapetian:2004tw}. If the thrust~\cite{farhi} 
of the quark jet can be empirically determined, then 
the Sivers asymmetry can be measured directly, where we replace 
$\mathbf{p}_\pi$ with $\mathbf{p}_q$~\cite{Brodsky:2002cx}. 
We consider the Sivers effect 
exclusively.

As discussed in Ref.~\cite{Brodsky:2002cx}, 
a non-zero Sivers SSA follows from the interference
 of two amplitudes ${\cal M}[\gamma^\ast p(J_p^y) \to F]$ of 
differing nucleon spin $J_p^y =\pm 1/2$ 
which couple to the same final-state $F$. In particular, the 
quantum numbers 
of the struck quark are the same in each case. 
The polarization axis $\mathbf{y}$
is chosen transverse to the scattering plane. 
The two amplitudes must also differ in their
 strong phase to generate a non-zero SSA, so that the SSA 
is proportional to Im$({\cal M}[J_p^y=+1/2]^\ast{\cal M}[J_p^y=-1/2])$. 
The photon cannot flip the helicity of the 
struck quark, so that 
the two amplitudes differ by $|\Delta L^y|=1$ to yield a non-zero 
result. The requisite matrix element is related, but not identical, to 
the matrix element which generates the anomalous magnetic 
moment~\cite{Brodsky:2002cx}. In particular, 
the presence of a strong phase engendered by FSI
is essential to generating a non-zero SSA. 

The Sivers SSA $A_{UT}^S$ is determined by the function 
$f_{1T}^{\perp\,q}(x,\mathbf{k}_\perp^2)$, which is subsumed
in $f_{q/p\,\uparrow} (x,\mathbf{k}_\perp)$, the 
distribution of unpolarized quarks in a transversely polarized
proton of spin $S$ and mass $M$; we define~\cite{Bacchetta:2004jz}
\begin{eqnarray}
f_{q/p\,\uparrow} (x,\mathbf{k}_\perp)
&=& f_1^q(x,\mathbf{k}_\perp^2) - 
f_{1T}^{\perp\,q}(x,\mathbf{k}_\perp^2) 
\frac
{ \epsilon^{\mu \nu \rho \sigma} P_{\mu} k_\nu S_\rho n_\sigma}{M (P\cdot n)}
\nonumber \\
&=& f_1^q(x,\mathbf{k}_\perp^2) - 
f_{1T}^{\perp\,q}(x,\mathbf{k}_\perp^2) 
\frac
{(\mathbf{\hat{P}}\times \mathbf{k}_\perp)\cdot \mathbf{S}}{M} \,,
\label{trento}
\end{eqnarray} 
in a frame where $\hat{P}$ and $n$, 
an auxiliary lightlike vector, point in opposite directions. 
We thus have 
\begin{equation}
A_{UT}^S = -\frac{2}{M} 
\frac{ \langle \sum_q  |\mathbf{k_\perp}| e_q^2
f_{1T}^{\perp\,q}(x,\mathbf{k}_\perp^2) 
{\cal D}(z,\mathbf{p}_\pi,\mathbf{k}_\perp)
\sin^2(\phi - \phi_S)
\rangle}{
\langle \sum_q e_q^2 f_1^q(x,\mathbf{k}_\perp^2) 
{\cal D}(z,\mathbf{p}_\pi,\mathbf{k}_\perp)
\rangle}\,,
\label{autsdef}
\end{equation} 
where 
${\cal D}(z,\mathbf{p}_\pi,\mathbf{k}_\perp)$ 
contains the fragmentation function 
$D_q^{\pi^{\pm}} (z,p_\pi)$
and 
$\langle \dots \rangle$ refer to the appropriate 
angle and $\mathbf{k_\perp}$ integrals~\cite{Anselmino:2005ea}. 
The quark electric charge $e_q$ dependence follows since  
$f_{q/p\,\uparrow} (x,\mathbf{k}_\perp)$ comes from the 
square of the scattering amplitude. 
The function $f_{1T}^{\perp\,q}(x,\mathbf{k}_\perp^2)$ can be extracted
from $f_{q/p\,\uparrow}$, which is defined in a gauge-invariant
way via~\cite{Belitsky:2002sm,Ji:2002aa}  
\begin{eqnarray}
&& \lefteqn{f_{q/p\,\uparrow} (x,\mathbf{k}_\perp)
= \int \frac{d\xi^- d^2 \bit{\xi}_\perp}{16 \pi^3} 
e^{-ik^+\xi^- + i \mathbf{k}_\perp\cdot \bit{\xi}_\perp}} \nonumber \\
&& \!\!\!\!\!\!\times\, 
\langle P
| \bar \psi (\xi_i, \bit{\xi}) 
[\infty, \bit{\infty}; \xi^-, \bit{\xi}_\perp]^\dagger_C 
\gamma^+ 
[\infty, \bit{\infty}; 0^-, \bit{0}_\perp]_C 
\psi(0,\bit{0}_\perp) | P
\rangle \,,
\label{jiyuandef}
\end{eqnarray} 
where $[\dots]_C$ denote gauge links stretched in both
light-like and transverse directions~\cite{Belitsky:2002sm}, 
capturing the final-state interactions necessary 
for the SSA. As we shall see, the latter 
imply that the Sivers
function can only be related, rather than identical to, 
the matrix element for the anomalous magnetic moment. 

The 
$\Delta L^y$ needed for a non-zero single-spin asymmetry can 
stem from two distinct physical sources: 
$\Delta L^y$ can arise from either quark {\em or} gluon degrees of 
freedom~\cite{Burkardt:2004ur,Anselmino:2004nk}. 
In the first case the virtual photon 
strikes a quark in a Fock component of the nucleon's light-front
wave function, whereas in the second case, the virtual photon fuses with
a gluon in some $| qqq g ...\rangle$ Fock state of the nucleon to 
produce a $q\bar q$ pair. 
Analogues of both mechanisms can also contribute 
to the nucleon's anomalous magnetic moment $\kappa$, as we shall
detail. However, the empirical anomalous magnetic moments of
the proton and neutron sum to nearly zero, suggesting that the
gluon contribution to the anomalous magnetic moment 
is small. We shall argue that a similar
cancellation observed in the SSA data from a deuteron target 
allows us to conclude that the gluon mechanism is small in this 
case as well. 
The two mechanisms we discuss are distinct, in part, because 
the virtual photon interacts with either an ``intrinsic'' quark, namely, 
a multiply-connected 
constituent of a multi-parton Fock state, or an ``extrinsic''
quark, produced as a member of a $q\bar q$ pair from photon-gluon fusion. 
The intrinsic quark carries an orbital angular momentum $L_q$, whereas
the extrinsic quark carries, in part, the orbital angular momentum
of the gluon constituent $L_g$ which spawns it. We can 
similarly distinguish
intrinsic from extrinsic gluons, so that an intrinsic gluon is also
a multiply-connected constituent of a multi-parton Fock state. 
Our separation can be clouded by 
QCD evolution effects, 
for an extrinsic gluon spawned by the DGLAP 
(Dokshitzer-Gribov-Lipatov-Altarelli-Parisi) 
splitting $q\to q g$ can fuse with the virtual photon to generate 
extrinsic quarks and a SSA.  
This mechanism thus serves to dilute the correlation 
between the gluon dynamics and the intrinsic gluon orbital 
angular momentum contribution to the proton spin. 

We emphasize that the two mechanisms, quark and gluon, 
differ in their isospin
character and, indeed, that SSA data on the proton and deuteron can
be used to distinguish them. We note that a large SSA for $\pi^+$
production from a transversely polarized proton has been observed 
by the HERMES collaboration~\cite{Airapetian:2004tw}; 
however, when an analogous observable 
is measured by the COMPASS collaboration 
from a deuterium target~\cite{Alexakhin:2005iw}, 
the SSA is consistent with {\em zero}. 
The polarization of the deuteron itself 
is used to define the spin correlation. 
The deuteron with spin $S_d^y = +1$ normal to the 
scattering plane has both nucleon spins aligned  
$S_p^y = +1/2$ and $S_n^y = +1/2$.  
Since the deuteron is a weakly bound state, the SSA from the
deuteron is the sum of single-spin asymmetries (SSAs) 
for the proton and neutron to 
a very good approximation. 
The small SSA observed with a $^2$H target 
is, in fact, natural if the matrix element is related to that 
 of the anomalous magnetic moment --- 
the empirical p and n anomalous magnetic moments sum to nearly zero. 
However, we shall show that this connection 
can only occur 
with the $L_q$ mechanism; 
the isospin structure of the $L_g$ mechanism is altogether distinct. 
In this regard whether the gluon-mediated SSA emerges 
from intrinsic or extrinsic gluons is without consequence. 
We note, in passing, that the use of other polarized nuclear targets
can give empirical checks of 
these observations; for example, 
polarized  $^3$He offers an 
effective neutron target, up to ``spin dilution'' corrections 
of some 10\%~\cite{ciofi}. 
We shall now develop these ideas in detail.

To set the stage, we review the manner
in which the anomalous magnetic moment of the nucleon is connected to 
the quark orbital angular momentum in the light-front formalism. 
Working in the 
interaction picture for the electromagnetic current 
$J^\mu(0)$ and the $q^+=0$ frame~\cite{DrellYan,West}, 
we have~\cite{BrodskyDrell} 
\begin{equation} 
\kappa  = - 
 \sum_{a} \sum_j e_j 
 \int
[{\mathrm d} x] [{\mathrm d}^2 \mathbf{k}_{\perp}] 
\psi^\ast_a(x_i,\mathbf{k}_{\perp i},\lambda_i)
\mathbf{S}_\perp\cdot \mathbf{L}_\perp^{q_j} 
\psi_a(x_i,\mathbf{k}_{\perp i},\lambda_i)
\equiv  \sum_{q} e_q a_q
 \,,
\label{kappadef}
\end{equation}
where we write $\kappa$ in units of $e/2M$ and 
define 
$\mathbf{S}_\perp\cdot \mathbf{L}_\perp^{q_j} 
\equiv (S_+ L_-^{q_j} 
+ S_- L_+^{q_j})/2$
with $S_\pm = S_1 \pm i S_2$ and 
$L_\pm^{q_j}
= \sum_{i\ne j} x_i ( 
{\partial}/{\partial k_{1i}} \mp i 
{\partial}/{\partial k_{2i}}) $ --- 
the last sum is over quark flavor $q$. 
Consequently, 
the orbital angular momentum contribution $L_\pm^{q_j}$ associated 
with a struck quark $j$ in Fock state $a$ is not an independent
variable, but, rather, is determined by the 
sum of the orbital angular momenta of {\em all} 
the spectator partons in that Fock state. This notion 
also gives rise to the vanishing anomalous {\it gravitomagnetic} moment
for composite systems, Fock state by Fock state~\cite{Brodsky:2000ii}.  
%Although we regard 
%$\mathbf{L}_\perp^{q_j}$ 
%as the transverse 
%orbital angular momentum associated with
%quark $j$, it is apparent that the transverse
%orbital angular momenta carried by gluon 
%spectators implicitly contributes to its definition. 
Although we regard $\mathbf{L}_\perp^{q_j}$
as the transverse orbital angular momentum associated with 
the struck quark $j$, the explicit sum over $i\ne j$
makes it apparent that the transverse orbital angular momenta carried by gluon
spectators implicitly contributes to its definition. Moreover, both quark
and gluon contributions from the parent nucleon Fock state are captured 
by the matrix element of the $L_\perp^{q_j}$ operator, as the 
gluon can fluctuate to a $q\bar q$ pair, to which the photon can couple. 
It is these quark- and gluon-mediated contributions 
%this last possibility 
which we distinguish as the ``quark''
and ``gluon'' mechanisms. We note that the light-front formalism
in $A^+=0$ gauge permits a simple kinematic 
operator representation of the $\mathbf{L}_z$ operator; this, in turn, permits 
$\mathbf{S}_\perp\cdot \mathbf{L}_\perp^{q_j}$ 
to act as a ladder operator which raises or lowers the value of 
$\mathbf{L}_z$ in this representation.
In the 
last equality of Eq.~(\ref{kappadef}) we subsume the Fock-state
sum to define the contribution to the anomalous magnetic moment,
quark flavor by quark flavor; the $a_j$ are real. 
The phase-space integration is given by 
\begin{equation}
\int [\mathrm{d}x]\,[\mathrm{d}^2\mathbf{k}_\perp] 
\equiv \!\!\sum_{\lambda_i, c_i, f_i}\!
\left[ \,
 \prod_{i=1}^{n}
\left( \int\!\!\int {{\rm d}x_i\, {\rm d}^2 {\mathbf{k}_{\perp i}}
\over  2 (2 \pi)^3}\ \right) \right] 
16\pi^3 \delta\left(1-\sum_{i=1}^{n} x_i\right) 
\delta^{(2)}\left(\sum_{i=1}^{n} {\mathbf{k}_{\perp i}}\right)\,, 
\label{phasespace}
\end{equation}
where $n$ denotes the number of constituents in Fock state $a$, 
and we sum over the possible $\{\lambda_i\}$, $\{c_i\}$, and $\{f_i\}$ 
in state $a$. 
The summations are over all contributing Fock states $a$ and struck
constituent charges $e_j$; we refrain from including 
the constituents' color and flavor dependence in the arguments of 
the light-front 
wave function (LFWF) $\psi^{S_z}_a$, which we define in the 
$A^+=0$ gauge, with the principal-value prescription for 
singularities in $k^+$. 
We emphasize that both quark and gluon degrees 
of freedom in the nucleon's LFWF 
contribute to Eq.~(\ref{kappadef}). 
Either an intrinsic or extrinsic 
gluon constituent in the nucleon 
%proton's LFWF
can couple to a photon via a $q\bar q$ pair. The 
lowest-order effective $\gamma^\ast g g$ vertex, a contribution
to the anomalous magnetic moment, 
 is forbidden by $C$ invariance, 
though the radiation of an extra gluon from the effective vertex 
removes this constraint and 
makes it finite. We note the gluon mechanism should contribute to $\kappa_p$ 
and $\kappa_n$ with equal weight, up to isospin-breaking
effects, yet the empirical isoscalar
magnetic moment of the nucleon,
$\kappa_S \equiv (\kappa_n + \kappa_p)/2 = -0.06$,
 is numerically very small relative to $|\kappa_{n,p}|$
--- suggesting 
that the gluon mechanism is itself small. 

Returning to the Sivers function, we define
\begin{equation}
f_{q/p\,\uparrow} (\eta,\mathbf{l}_\perp) = 
\bar u(P,\lambda^\prime) \left[ 
f_{1}^{q}(\eta,\mathbf{l}_\perp^2) \gamma^+ - 
f_{1T}^{\perp\,q}(\eta,\mathbf{l}_\perp^2) i \sigma^{+\alpha} 
\frac{\mathbf{l}_{\perp\,\alpha}}{M}
\right] u(P,\lambda) \,,
\end{equation} 
where $u(P,\lambda)$ is a Dirac spinor associated with a spin-$1/2$ 
state of momentum $P$ and helicity $\lambda$~\cite{Lepage:1980fj}, 
with $\mathbf{y}$
the polarization axis --- only the $\alpha=1,3$ matrix elements are
nonzero. We identify $f_{q/p\,\uparrow} (\eta,\mathbf{l}_\perp)$ 
as the function $q(\eta,\mathbf{l}_\perp)$ of Ref.~\cite{Belitsky:2002sm}, 
where we work in leading twist and ignore all QCD evolution effects. 
Turning to the light-front formalism in $A^+=0$ gauge, with boundary
conditions appropriate to SIDIS, 
a Fock component of the proton's
LFWF has the form 
$\tilde\psi^{S_y}_a = \psi^{S_y}_a \exp(i\phi_a^{S_y})$; we emphasize
that the LFWFs are complex in this case~\cite{Belitsky:2002sm}. 
Note that we contrast the LFWF for a proton in isolation to that
for a proton immersed in an external electromagnetic gauge field. 
For simplicity we assume the LFWFs differ only in the phase
$\phi_a^{S_y}$. 
Working in the $q^+=0$ frame we thus identify 
\begin{eqnarray}
&&\!\! f_{1T}^{\perp\,q}(\eta,\mathbf{l}_\perp^2) 
\frac{l_{1}}{M}
= -\frac{i}{2} \sum_a \sum_{j=1}^n \delta_{q q_j} 
\int [dx] [d^2 \mathbf{k}_\perp] \Bigg\{ 
\psi_a^{\uparrow\,\ast}(x_i^\prime, \mathbf{k}_{\perp i}^\prime, \lambda_i) 
\psi_a^{\downarrow}(x_i, \mathbf{k}_{\perp i}, \lambda_i) 
\nonumber \\
&& \!\!\!\!\!\!\!\!\!\times \, 
\hbox{Im}\,\exp\left({i (\phi^{\downarrow} -\phi^{\uparrow})}\right)
+ 
\psi_a^{\downarrow\,\ast}(x_i^\prime, \mathbf{k}_{\perp i}^\prime, \lambda_i) 
\psi_a^{\uparrow}(x_i, \mathbf{k}_{\perp i}, \lambda_i) 
\hbox{Im}\, 
\exp\left({-i(\phi^{\downarrow} -\phi^{\uparrow})}\right) \Bigg\} \,,
\label{ftqLF} 
\end{eqnarray} 
where $\mathbf{k}'_{\perp j}=\mathbf{k}_{\perp j}+ \mathbf{l}_{\perp}$
and $x_j' = x_j + \eta $
for the struck constituent $j$ and
$\mathbf{k}'_{\perp i}=\mathbf{k}_{\perp i}-x_i \mathbf{l}_{\perp}/(1 - x_j)$
and $x_i' = x_i[1 - \eta/(1-x_j)] $
for each spectator $i$, where $i\ne j$. 
The existence of a term in $\mathbf{l_\perp}$ mandates 
not only orbital angular momentum~\cite{Brodsky:2003pw} 
but also 
the imaginary parts in the right-hand side (RHS) of Eq.~(\ref{ftqLF}) 
--- a FSI phase must be present to incur a SSA. We have 
suppressed the arguments of $\phi^{S_y}$ but assert that 
it depends on the magnitude and not the direction of 
$\mathbf{k}_{\perp i}^{(\prime)}$, 
so that 
the $|\Delta L^y|=1$ structure of the matrix element entails the 
${l}_{1}$ dependence, specifically that the RHS 
$\sim -(i/2)[(l_{3} -i l_{1} ) - (l_{3} +i l_{1} )]
=- l_{1}$, 
as found in explicit model 
calculations~\cite{Brodsky:2002cx,Ji:2002aa,signnote}. The Kronecker $\delta$
ensures that the struck quark is of flavor $q$. 
Following 
the development of Eq.~(\ref{kappadef}) in Ref.~\cite{BrodskyDrell}, we 
find that $f_{1T}^{\perp\,q}(x,\mathbf{l}_\perp^2)$ can also be written
in terms of the matrix element of a spin-orbit operator, 
if we consider
the leading terms as $\mathbf{l}_\perp \to 0$: 
\begin{eqnarray}
\frac{f_{1T}^{\perp\,q}(\eta,0)}{M}
\!\!&=& \!\!
\sum_{a} \sum_j \delta_{q q_j}
 \int
[{\mathrm d} x] [{\mathrm d}^2 \mathbf{k}_{\perp}] 
\frac{1}{2i(1-x_j)} 
\Bigg[
\tilde \psi^\ast_a(x_i^\prime, \mathbf{k}_{\perp i}, \lambda_i) 
\mathbf{\bar S}_{\perp T}\cdot \mathbf{\bar L}_{\perp T}^{q_j} 
\tilde \psi_a  (x_i, \mathbf{k}_{\perp i}, \lambda_i) 
\nonumber \\
&& - 
\bar \psi^\ast_a(x_i^\prime, \mathbf{k}_{\perp i}, \lambda_i) 
\mathbf{\bar S}_{\perp T}\cdot \mathbf{\bar L}_{\perp T}^{q_j} 
\bar 
\psi_a  (x_i, \mathbf{k}_{\perp i}, \lambda_i) \Bigg] \,,
\label{atilde1}
\end{eqnarray}
where
$\bar\psi^{S_y}_a \equiv \psi^{S_y}_a \exp(-i\phi_a^{S_y})$, 
$\mathbf{\bar S}_{\perp T}\cdot \mathbf{\bar L}_{\perp T}^{q_j} 
\equiv (S_{+T} L_{-T}^{q_j} 
- S_{-T} L_{+T}^{q_j})/2$, 
$S_{\pm T} = S_3 \pm i S_1$, and 
$L_{\pm T}^{q_j}
= \sum_{i\ne j} x_i ( 
{\partial}/{\partial k_{3i}} \mp i 
{\partial}/{\partial k_{1i}})$. We define 
\begin{equation}
\frac{f_{1T}^{\perp\,q}(\eta,\mathbf{l}_\perp^2)}{M}
\equiv  - \tilde a_q (\eta, \mathbf{l}_\perp^2)\,, 
\label{atildedef}
\end{equation}
where the $\tilde a_j$ are real. 
A comparison of 
Eqs.~(\ref{kappadef}) and (\ref{atilde1}) prompts us
to include a minus sign in the definition of 
$\tilde a_q (x, \mathbf{l}_\perp^2)$. 
We note in passing that the 
generalized parton distribution $E(x,\zeta,t)$ probed in 
virtual Compton scattering (VCS) 
can also be connected to the nucleon's orbital angular momentum. 
The generalized form factors in VCS, 
$\gamma^\ast(q) + p(P) \to \gamma^\ast(q^\prime) + p(P^\prime)$
with $t=\Delta^2$ and 
$\Delta = P - P^\prime = (\zeta P^+, 
\mathbf{\Delta}_\perp, (t + \mathbf{\Delta}_\perp^2)/\zeta P^+)
$, have been constructed 
in the light-front formalism~\cite{Brodsky:2000xy}. 
Using Eq.~(40) of Ref.~\cite{Brodsky:2000xy}, and 
the procedure and syntax 
of Eq.~(\ref{kappadef}), noting $\mathbf{q}_\perp \to \mathbf{\Delta}_\perp$, 
we determine, for $\zeta\le x \le 1$, 
%if we consider the leading terms as $\mathbf{\Delta}_\perp \to 0$: 
\begin{eqnarray}
%\begin{equation}
\frac{E(x,\zeta,0)}{2M} &=& \sum_a 
(\sqrt{1 -\zeta})^{1-n} \sum_j \delta(x -x_j) 
\int [{\mathrm d} x] [{\mathrm d}^2 \mathbf{k}_{\perp}] 
\quad\quad\quad\nonumber \\
&&\times \,
\psi^\ast_a (x_i^\prime ,\mathbf{k}_{\perp i},\lambda_i)
\mathbf{S_\perp}\cdot \mathbf{L_\perp^{q_j}} 
\psi_a (x_i,\mathbf{k}_{\perp i},\lambda_i) \,, 
\end{eqnarray}
%\end{equation}
with $x_j^\prime=(x_j -\zeta)/(1-\zeta)$ for the struck parton $j$
and $x_i^\prime=x_i/(1-\zeta)$ for the spectator parton $i$. 
We emphasize that the LFWFs for Fock-state $a$ 
with spin up or down 
for fixed struck quark helicity differ by $|\Delta L^z|=1$
because $\mathbf{L}_\perp^{q_j}$ contains ladder operators. 

\section*{The SSAs and the Anomalous Magnetic Moments}

We see 
that the matrix 
element, Eq.~(\ref{atilde1}), which drives the Sivers SSA 
bears similarity 
to that of the 
anomalous magnetic moment~\cite{Brodsky:2002cx,Burkardt:2002ks,Burkardt:2003uw,Burkardt:2003je,Burkardt:2003yg,Burkardt:2005hp,Burkardt:2005km}.
To understand the consequences of this in a transparent way, 
we recall that under an assumption of isospin symmetry, we have 
$a^p_d = a^n_u$, $a^p_u = a^n_d$,
$a^p_{\bar d} = a^n_{\bar u}$, $a^p_{\bar u} = a^n_{\bar d}$, and 
$a^p_{\bar q} = a^n_{\bar q}$ for sea quarks of other flavors. 
Isospin symmetry acts at the level of the hadronic matrix elements; 
the quark charges are not isospin mirrors. 
Neglecting the contributions of anti-quarks we have 
\begin{eqnarray}
&& \kappa_p=1.79 = (+2/3)a^p_u + (-1/3)a^p_d \nonumber \\
&&\kappa_n = -1.91 = (-1/3)a^n_d + (+2/3)a^n_u  
= (+2/3) a^p_d + (-1/3) a^p_u  \,, 
\end{eqnarray} 
to yield 
 $a^p_d = a^n_u = -2.03$ and  $a^p_u = a^n_d = 1.67$. 
Concommitant isospin relations follow for the $\tilde a_q$ of 
Eq.~(\ref{atildedef}) as well. 
We also neglect the contributions of the anti-quarks to the 
existing SSA data, which is consistent with recent 
fits~\cite{Anselmino:2005ea,Vogelsang:2005cs,naomi}. 
In what follows, if we {\it conjecture}
that the 
isospin structure of the empirical anomalous magnetic moments is that
of the Sivers SSA, then we find 
the relative strength of the various 
$\tilde a_q^p(\eta,\mathbf{l}_\perp^2)$ can be 
estimated through that of the 
$a_q^p$. Recalling Eqs.~(\ref{autsdef}) and (\ref{atildedef}), 
 the negative sign of $a^p_d$ predicts a negative
 sign of the SSA for $l p \to l^\prime \pi^- X$, whereas 
 $a^p_u = 0.835$ predicts a positive asymmetry for mesons produced by
favored fragmentation from the $u$ quark. 
As noted in Refs.~\cite{Burkardt:2003uw,Burkardt:2003je,Burkardt:2003yg}, 
both predictions are consistent with HERMES data 
at sufficiently large $z$~\cite{Airapetian:2004tw}. 

As to the magnitudes of the 
SSAs, although $|a^p_u| < |a^p_d|$, 
the SSA engendered by $u$-quark
fragmentation from a proton is enhanced by a factor of $4$,
since the asymmetry is controlled by $e_u^2$. 
In the absence of anti-quark contributions, 
the SSA for $\pi^-$ production should indeed be small, and this is
observed~\cite{Airapetian:2004tw}. 
Recent model fits~\cite{Anselmino:2005ea,Vogelsang:2005cs}
are consistent with these
trends. To be more specific, we note the fits of 
Ref.~\cite{Vogelsang:2005cs} yield 
$S_u=-0.81\pm 0.07$ and $S_d=1.86\pm 0.28$, where 
$S_q$ is to be compared to $-\langle\tilde a_q\rangle$, 
with $\langle\tilde a_q\rangle$ defined to be the average value of 
$\tilde a_q (\eta, \mathbf{l}_\perp^2)$. 
This implies that the SSA asymmetry from $u$-quark fragmentation
is smaller than that predicated from the use of 
the empirical anomalous magnetic moments alone. 
Although we expect the strong phase from the Wilson line
to be isoscalar, 
the extra factor of  $1/(1- x_j)$ in the matrix element
of Eq.~(\ref{atilde1}) could change the relative strength
of the $u$ and $d$ contributions. 
Nevertheless, 
considering the consequences of this simple picture
for the deuteron data, we note that 
$a^p_u + a^n_u = a^p_d + a^n_d = -0.360$, implying 
that the SSA for $u$-quark fragmentation to leading positively
charged hadrons, as well as $d$-quark fragmentation to leading 
negatively charged hadrons, ought be small. 
This is borne out by the recent COMPASS data 
in SIDIS from a deuteron target~\cite{Alexakhin:2005iw}. Such a cancellation is
consequent to the differing signs of 
$a_u^{p,n}$ and $a_d^{p,n}$~\cite{Anselmino:2005ea,Vogelsang:2005cs}.

\section*{Quark or Gluon Orbital Angular Momentum?}

We now wish to use the differing isospin structure of the $L_q$ and
$L_g$ mechanisms to infer the relative size of these contributions 
to the total orbital angular momentum of the nucleon. 
The isospin structure of the $L_g$ mechanism is distinctive, for
a gluon in a nucleon Fock state will produce $u\bar u$ or $d\bar d$ pairs 
with equal weight --- up 
to tiny differences driven by the $u$-$d$ mass difference. Thus the $u$-quark
and $d$-quark SSAs add constructively in SIDIS from a $^2$H target; 
there is no cancellation of this $I=0$ physics. 
However, the $L_g$ mechanism cannot always generate a leading
hadron, i.e., a contribution which survives in the $z\to 1$ limit. 
For example, 
to realize the $l p \to l^\prime \pi^\pm X$ reaction via the  $L_g$ mechanism, 
{\em two} $q\bar q$ pairs must be produced to recover a $\pi^\pm$ hadron. 
It is significant, then, that the SSAs from the deuteron are 
consistent with zero for $z > 0.35$~\cite{Alexakhin:2005iw}, for 
both positively and negatively charged leading 
hadrons~\cite{leading}. 
In this kinematic region 
the SSA asymmetry in $\pi^+$ production on the proton is increasing,
so that the gluon mechanism can contribute in this $z$ region.
Moreover, the value of $x$, for leading hadrons, 
ranges from $0.006$ to $0.3$, so that 
the gluon mechanism can contribute in this $x$ region. 
The empirical SSAs in leading hadrons from $^2$H are 
consistent with zero, though, for all but the smallest $z$. 
In the context of the $L_q$ mechanism this can be understood as
emerging from an approximate cancellation of the $p$ and $n$ SSAs,
reflective of the isospin structure of the anomalous 
magnetic moments.  
The $^2$H data thus 
allow us to conclude that $L_g$ is small compared to the quark contributions. 
We note in passing that theoretical arguments based on the 
large $N_c$ expansion lead to a similar conclusion~\cite{metz}.

We can also crudely quantify 
the extent to which the $L_g$ mechanism is absent. 
 To estimate the relative size of the strong phase in the 
$L_q$ and $L_g$ cases, 
 we employ the same reasoning as used in interpreting 
the ratio of the rapidity plateaus in gluon versus 
quark jets~\cite{Brodsky:1976mg,Konishi:1978yx,gary}.
That is, we assume 
the phases scale, gluon to quark, as $2.25$, 
as per a leading-order analysis in 
the QCD coupling~\cite{Brodsky:1976mg,Konishi:1978yx,gary}.
Then, 
if we compute the 
ratio of the SSA from positively charged, leading hadrons on a 
$^2$H target~\cite{Alexakhin:2005iw} 
to that from leading $\pi^+$ production on a 
proton target~\cite{Airapetian:2004tw}, and
divide by a factor 
of roughly $2\cdot 2.25\approx 4.5$, as the deuteron contains
two nucleons, we can estimate the relative strength of the two
mechanisms. We find that the gluon mechanism is smaller than
the quark mechanism by a factor of $0.2$. 
The nucleon's
orbital angular momentum appears to be largely carried by its quarks. 
We can also use the ratio of SSAs for $u$- 
to $d$-quark jets in SIDIS on the proton 
to determine the ratio of 
$\tilde a^p_u/ \tilde a^p_d$ as a function of $\eta$ and 
$\mathbf{l}_\perp^2$. 
To realize this experimentally, 
one needs to work at large $z$ where the jet tagging is
reliable, 
 i.e., where the hadron type tags the flavor of the ``struck'' quark.

We have argued that the $L_g$ mechanism is small and cannot
always produce a leading hadron, so that one is left to ponder how current
empirical constraints 
can be bettered. It strikes us
as efficacious to study SSAs 
associated with produced hadrons
of non-valence quark content. 
The $\gamma^* g \to s \bar s \to K^-K^+ + X$ reaction is one such possibility. 
In principle, 
one can trace the SSA of the $K^-K^+$ to the 
gluon's orbital angular momentum $L_g$.
One can also consider the 
$\gamma^* g \to s \bar s \to \phi + X$ reaction: the SSA in $\phi$ production. 
Both reactions are 
important tests for the $L_g$ mechanism, since 
the gluon contributions of the two nucleons to the SSA
add. One can consider these processes as aspects of the gluon jet.  
In this, we ignore the 
possibility of intrinsic strangeness in the nucleon's 
non-perturbative structure, since 
parity-violating electron scattering experiments
show the strangeness contribution to the proton's anomalous magnetic moment 
to be small~\cite{parity}. 
We note that Anselmino et al. 
have discussed accessing the Sivers gluon distribution
through open charm production~\cite{Anselmino:2004nk}; 
this is similar in conception  
to the suggestions we offer here. 

The Sivers asymmetry from gluons can also be studied directly, if 
the 
empirical thrust~\cite{farhi} of the gluon jet axis can be
determined. If this
could be done, the study of the $L_g$ mechanism would be much
facilitated, as the Collins mechanism would no longer contribute. 
To extract detailed numerical information about the gluon 
mechanism from the SSAs, one 
ultimately requires information about the strong phase from FSI; nevertheless, 
the experiments we suggest do serve to bound the size of the gluon's orbital
angular momentum contribution to the nucleon's spin. 

Let us conclude with a brief summary. 

\begin{itemize}

\item   A non-zero SSA follows from the interference
 of two amplitudes of differing nucleon spin, 
but of common quark helicity.  

\item  The two amplitudes must also differ in their
  strong phase to generate a non-zero SSA, so that
the matrix element which yields the Sivers function
cannot be identical to that for the anomalous magnetic
moment. Nevertheless, the signs of the SSAs from 
leading $u$- and $d$-quark fragmentation from the proton 
correlate with those of the $u$ and $d$-quark contributions
to the anomalous magnetic moment, analyzed under an
assumption of isospin symmetry. 

  \item The anomalous magnetic moment, the Sivers function, 
and the generalized parton distribution $E$ can all be
connected to matrix elements involving the orbital 
angular momentum of the nucleon's constituents. 
To be specific, the matrix elements are between 
LFWFs that differ
by one unit of orbital angular momentum along the polarization
axis; 
this follows as we compute matrix elements of the angular
momentum lowering or raising operator.

 \item The SSA can be generated by either a quark or
   gluon mechanism, and the isospin structure of the
   two mechanisms is distinct. The approximate
   cancellation of the SSA measured on a deuterium target 
   suggests that the gluon mechanism, and thus the orbital angular
   momentums carried by gluons in the nucleon, is small.

\end{itemize}

\smallskip
\noindent{\bf Acknowledgments:}\,\,
S.G. thanks Wolfgang Korsch for bringing the single spin asymmetry
data on $^2$H to her attention and the SLAC theory group for
gracious hospitality.

{\noindent{\it Note added:} After the completion of our paper,} 
we became aware of 
work contemporary to ours by Anselmino {\it et al.}, 
Ref.~\cite{Anselmino:2006yq}, which 
draws similar conclusions from the measured 
SSA in $p^\uparrow p \to \pi^0 X$ scattering.


\begin{thebibliography}{99}
\bibitem{leading} In this study~\cite{Alexakhin:2005iw}
the COMPASS collaboration defines
the leading hadron as the most energetic hadron with 
$z>0.25$ which originates from the reaction vertex.
\end{thebibliography}
\end{document}